\documentclass[onecolumn,floatfix,showpacs,tightenlines,superscriptaddress]{revtex4-2}
\usepackage{mathtools}
\usepackage{bm}
\usepackage{dsfont,amsthm,amsbsy}
\usepackage{verbatim}
\usepackage{amssymb}
\usepackage{amsmath}
\usepackage{bbm}
\usepackage{graphicx}
\usepackage{epstopdf}
\usepackage{subfigure}
\usepackage{natbib}
\usepackage{epsfig}
\usepackage{amsfonts}
\usepackage{mathrsfs}
\usepackage{sidecap}
\usepackage{lipsum}
\usepackage[toc,page,title,titletoc,header]{appendix}
\usepackage[colorlinks,linkcolor=blue,citecolor=blue,anchorcolor=blue,urlcolor=blue]{hyperref}
\usepackage{hyperref}
\usepackage{resizegather}
\usepackage{tikz}
\usepackage{float}
\usepackage{mathbbol}
\usepackage[normalem]{ulem}
\usepackage{cancel}
\usepackage{upgreek}

\newcommand{\bl}{\begin{aligned}}
\newcommand{\el}{\end{aligned}}
\def\be{\begin{equation}}
\def\ee{\end{equation}}

\def\bi{\begin{itemize}}
\def\ei{\end{itemize}}
\def\bn{\begin{enumerate}}
\def\en{\end{enumerate}}
\def\bea{\begin{eqnarray}}
\def\eea{\end{eqnarray}}
\def\no{\nonumber}
\def\ba{\begin{array}}
\def\ea{\end{array}}
\def\bd{\begin{displaymath}}
\def\ed{\end{displaymath}}

\begin{document}

\title{Scaling and Universality at Ramped Quench Dynamical Quantum Phase Transitions}

\author{Sara Zamani}
\affiliation{Department of Physics, Institute for Advanced Studies in Basic Sciences (IASBS), Zanjan 45137-66731, Iran}

\author{J. Naji}
\affiliation{Department of Physics, Faculty of Science, Ilam University, Ilam, Iran}

\author{R. Jafari}
\email[]{jafari@iasbs.ac.ir, raadmehr.jafari@gmail.com}
\affiliation{Department of Physics, Institute for Advanced Studies in Basic Sciences (IASBS), Zanjan 45137-66731, Iran}
\affiliation{School of Nano Science, Institute for Research in Fundamental Sciences (IPM), 19395-5531, Tehran, Iran}
\affiliation{Department of Physics, University of Gothenburg, SE 412 96 Gothenburg, Sweden}

\author{A. Langari}
\email[]{langari@sharif.edu}
\affiliation{Department of Physics, Sharif University of Technology, P.O.Box 11155-9161, Tehran, Iran}

\begin{abstract}
The nonequilibrium dynamics of a periodically driven extended XY model, in the presence of linear time dependent magnetic filed, is
investigated using the notion of dynamical quantum phase transitions (DQPTs). 
Along the similar lines to the equilibrium phase transition, the main purpose of this work is to search the fundamental concepts
such as scaling and universality at the ramped quench DQPTs.
We have shown that the critical points of the model, where the gap closing occurs, can be moved by tuning the driven frequency and consequently the presence/absence
of DQPTs can be flexibly controlled by adjusting the driven frequency.
We have uncovered that, for a ramp across the single quantum critical point, the critical mode at which DQPTs occur is classified into three regions:
the Kibble-Zurek (KZ) region, where the critical mode scales linearly with the square root of the sweep velocity, pre-saturated (PS) region, and the saturated (S) region
where the critical mode makes a plateau versus the sweep velocity. While for a ramp that crosses two critical points, the critical modes disclose just KZ and PS regions.
On the basis of numerical simulations, we find that the dynamical free energy scales linerly with time, as approaches to DQPT time, with the exponent $\nu=1\pm 0.01$  for all sweep velocities and driven frequencies.
\end{abstract}

\maketitle

\section{Introduction}
In the recent decades remarkable attention has focused on studying the out-of-equilibrium physics
of low dimensional quantum systems
\cite{Polkovnikov2011,Cazalilla2011,Bernard2016,Gogolin2016,Abanin2019,Nava2023,Nava2023b}.
The renaissance of the topic was commenced by the experimental advances achieved with ultra-cold atoms in optical lattices \cite{jotzu2014experimental,daley2012measuring,schreiber2015observation,flaschner2018observation}
which made possible to prepare and control non-equilibrium states with previously unexpected controllability
and stability \cite{BlochRevModPhys2008}.
Thereafter, trapped ions \cite{jurcevic2017direct,martinez2016real,smith2016many}, nitrogen-vacancy centres in diamond \cite{yang2019floquet}, superconducting qubit systems \cite{guo2019observation} and quantum walks in photonic systems \cite{wang2019simulating,xu2020measuring} developed to provide a framework for studying experimentally a wide variety of non-equilibrium systems.
These experiments have also provoked huge progress in theoretical physics.\\

Recently, new research direction of quantum phase transition proposed theoretically in nonequilibrium quantum systems,
titled dynamical quantum phase transitions (DQPTs) \cite{Heyl2013,Heyl2018} as a counterpart of equilibrium
phase transitions. The concept of DQPT emanates from the analogy between the equilibrium partition function of a system, and Loschmidt amplitude,
which measures the overlap between an initial state and its time-evolved one
\cite{Heyl2013,Heyl2018,Jafari2019quench,Najafi2018,Jafari2017,Najafi2019,Yan2020,Zache2019,Mukherjee2019,Zhang2016,Zhang2016b,Serbyn2017,
Sadrzadeh2021,Cao2022,Wong2022,Sehrawat2021,Rylands2021,Kloc2021,Bhattacharyya2020,Abdi2019,Knaute2023}.
As the equilibrium phase transition is signaled by non-analyticities in the thermal free energy, the DQPT is revealed
through the nonanalytical behavior of dynamical free energy, where the real-time plays the role of the control parameter \cite{andraschko2014dynamical,vajna2015topological,Karrasch2013,vajna2014disentangling,jafari2019dynamical,Mondal2022,Mendoza2022,
Sedlmayr2018,Sedlmayr2018b,Khatun2019,Teemu2021,Ding2020,Corps2023,Nicola2022,Nicola2021,Verga2023,Rossi2022,Khan2023anomalous,Khan2023,Cheraghi2023}.
DQPT displays a phase transition between dynamically emerging quantum phases, that takes place during the nonequilibrium
coherent quantum time evolution under sudden/ramped quench \cite{zhou2018dynamical,Jafari2020,Zhou2021,Mishra2020,Vanhala2023,Mondal2023,Hoyos2022,Cao2020,Sedlmayr2023,Sedlmayr2022,Corps2022,Sedlmayr2020,
Hou2022,Modak2021,Zeng2023,Stumper2022,Yu2021,Vijayan2023,Xue2023,Bhattacharjee2023,Leela2022,Porta2020,Puskarov2016,Karrasch2017} or time-periodic
modulation of Hamiltonian \cite{yang2019floquet,Zamani2020,kosior2018dynamical,Jafari2021,kosior2018dynamicalb,Naji2022,Jafari2022,Naji2022b}.
In addition, a dynamical topological order parameter (DTOP) has been proposed to capture DQPTs \cite{budich2016dynamical,Sim2022},
analogous to order parameters at equilibrium quantum phase transition.
DTOP reveals integer values as a function of time and its unit magnitude jumps, at the dynamical phase transition times,
manifest the topological distinctive feature of DQPTs \cite{Bhattacharjee2018,Dutta2017,sharma2014loschmidt}.
Just a while ago, DQPT was observed experimentally in several studies \cite{flaschner2018observation,jurcevic2017direct,martinez2016real,guo2019observation,wang2019simulating,Nie2020,Tian2020,Francisco2022,Shen2023}
to confirm theoretical anticipation. Most of these researches associated with quantum evolution
generated by a sudden quench of the Hamiltonian. However, comparatively little attention has been devoted to the
more realistically driving systems, ramp quench, and specifically, the scaling and universality at ramped quench DQPTs.
Understanding the scaling and universality at ramped quench DQPTs is of utmost importance both
in designing experiments and comprehend the results \cite{Marino2012,Marino2014}.
In this paper, we try to contribute to understand the scaling and universality of the non-equilibrium properties of the model
using the notion of the DQPTs. Using the advantage of controlling the location of the equilibrium phase transition points by driven frequency,
we numerically show that the  behaviour of the critical momentum at which the DQPTs occur can be categorize into three distinct regions versus the sweep velocity.
Moreover, we find that the dynamical free energy close to DQPT time shows a power law scaling with the exponent $\nu=1\pm0.01$, for any sweep velocities which is the same as that of the sudden quench.

The paper is organized as follows. In Sec. \ref{DPT}, the dynamical free energy and DTOP of the two band Hamiltonians are discussed.
In Sec. \ref{model} we present the model and review its exact solution and equilibrium phase transition.
Section \ref{results} is dedicated to the numerical simulation of the model based on the analytical result.
Section \ref{SD} contains some concluding remarks.

\section{Quench of an integrable model and dynamical phase transition\label{DPT}}

\subsection{Dynamical free energy\label{DFE}}
%
\begin{figure}
\centerline{\includegraphics[width=0.5\linewidth]{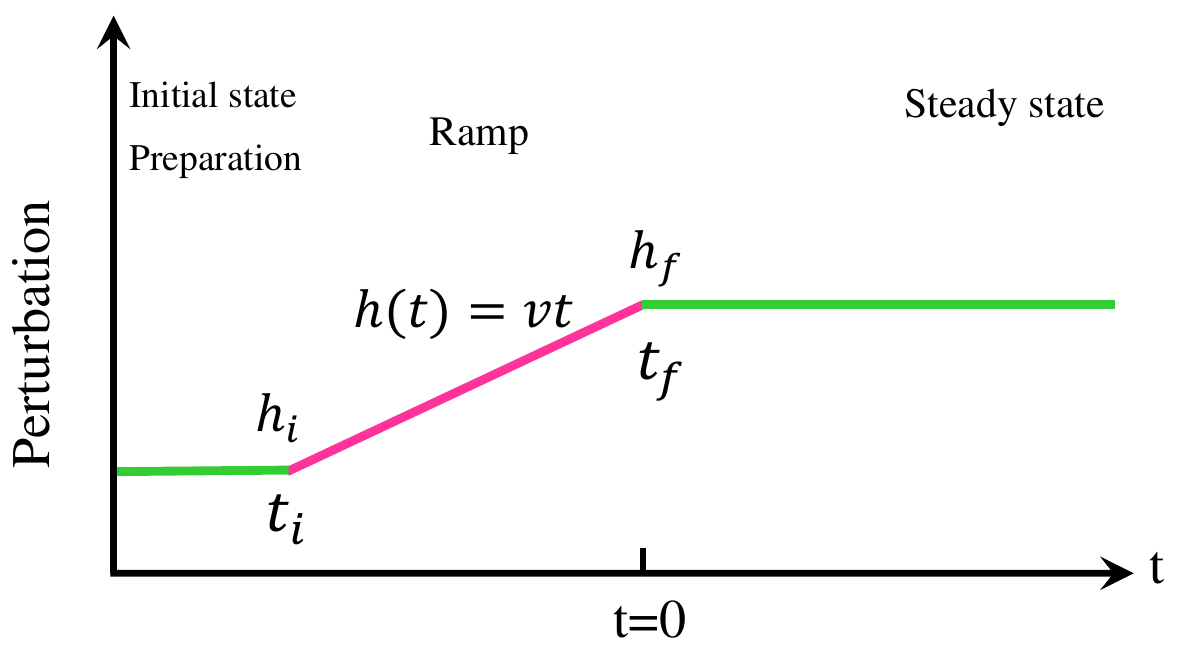}}
\caption{(Color online) Illustration of a linear ramped quench (pink color).
Here $h(t)$ is the magnetic field, $h_i$ and $h_f$ its initial and final values, 
and $t_i$ and $t_f=0$ the corresponding times.}
\label{fig1}
\end{figure}
%
For all ramped quench schemes, to be analyzed in the subsequent discussions, we follow the terminology implemented in Refs. \cite{Divakaran2016,Sharma2016b}.
Let us consider an integrable model reducible to a two level Hamiltonian $H_k(\lambda)$ for each momentum mode and
the system is initially ($t_i \to-\infty$) in the ground state $|1^i_k\rangle$  of the initial Hamiltonian $H_k(\lambda_i)$ for each mode.
In the ramped quench protocol, the Hamiltonian is characterized by a  parameter $\lambda$ which
is quenched from an initial value $\lambda_i$  at $t_i$ following the quenching protocol $\lambda(t)=vt$ to a final value $\lambda_f$ at $t_f$, in such a way
that the system crosses the QCP at $\lambda = \lambda_c$. Since the condition for an adiabatic dynamics breaks
in the vicinity of the QCP, one arrives at a final state $|\psi^{f}_{k} \rangle$ (for the $k$-th mode) which may not
be the ground state of the final Hamiltonian $H_{k}(\lambda_f)=H^{f}_{k}$. The final state can be written in the form of $|\psi^{f}_{k} \rangle = v_k |1_k^f\rangle+ u_k |2_k^f\rangle$, ($|u_k|^2 + |v_k|^2 =1$) where, $|1_k^f\rangle$ and $|2_k^f\rangle$ are the ground and the excited states of the final Hamiltonian $H^{f}_{k}$ with the corresponding energy eigenvalues {$\epsilon_{k,1}^f$ and $\epsilon_{k,2}^f$}, respectively.
Clearly $p_k=|u_k|^2=|\langle2_k^f|1^i_k\rangle|^2$ denotes the non-adiabatic transition probability that the system ends up at the excited state at the end of quench.
Therefore, the Loschmidt overlap for the mode $k$ for $t>t_f$ is defined by \cite{Divakaran2016,Sharma2016b}
%
\bea
\label{eq1}
{\cal L}_k=\langle \psi^{f}_{k} |\exp(-H^{f}_{k} t)| \psi^{f}_{k}\rangle
=|v_k|^2\exp(-i \epsilon_{k,1}^f t) + |u_k|^2 \exp(-i \epsilon_{k,2}^f t),
\eea
%
and the corresponding dynamical free energy \cite{Heyl2013,Heyl2018},  $g_k(t) = - \log \langle \psi^{f}_{k} |\exp(-H^{f}_{k} t)| \psi^{f}_{k} \rangle/N$,
where $N$ is the size of the system.\\

Summing over the contributions from all momentum modes and converting summation to the integral in the thermodynamic limit, one gets \cite{Divakaran2016,Sharma2016b,Pollmann2013}
%
\bea
\no
g(t)=\frac{-1}{2\pi} \int_{0}^{\pi} \log \Big(1 + 4 p_k (p_k-1) \sin^2 (\frac{\epsilon_{k,2}^f-\epsilon_{k,1}^f}{2}) t \Big) dk,
\eea
where $t$ is measured from the instant the final state, $|\psi^{f}_{k}\rangle$, is reached
at the end of the ramped quench (Fig. \ref{fig1}).
The argument of the logarithm is seen to vanish with $g(t)$ becoming nonanalytic when
$t=t_n^*$ given by
%
\bea
t_n^{\ast} =  \frac{\pi} {(\epsilon_{k^{\ast},2}^f-\epsilon_{k^{\ast},1}^f) }  \left(2n+ 1 \right).
\label{eq3}
\eea
%
These are the critical times for the DQPTs, with $k^{\ast}$ the mode that satisfies $|u_{k^{\ast}}|^2=p_{k^{\ast}}=1/2$.
For the case $\epsilon_{k,2}^f = -\epsilon_{k,1}^f  = \epsilon_k^f$, Eq. \eqref{eq3} is simplified to
%
\bea
t_n^{\ast} = \frac{\pi} {\epsilon_{k^{\ast}}^f}  \left(n+\frac 1 {2}\right).
\label{eq4}
\eea
%

\subsection{Dynamical Topological Order Parameter\label{DTOP}}

Analogous to order parameters at equilibrium quantum phase transition, a dynamical topological order parameter
is introduced to capture DQPTs \cite{budich2016dynamical,bhattacharya2017emergent}.
The DTOP is quantized and its unit magnitude jumps at the time of DQPTs reveal the topological aspect of DQPT \cite{budich2016dynamical,Bhattacharjee2018,sharma2014loschmidt}.
This DTOP is extracted from the guage-invariant Pancharatnam geometric
phase associated with the Loschmidt amplitude \cite{budich2016dynamical,Bhattacharya}.
The dynamical topological order parameter is defined as \cite{budich2016dynamical}
%
\begin{eqnarray}
\label{eq5}
W_n(t)=\frac{1}{2\pi}\int_0^\pi\frac{\partial\phi^G(k,t)}{\partial k}\mathrm{d}k,
\end{eqnarray}
%
where the geometric phase $\phi^G(k,t)$ is gained from the total phase $\phi(k,t)$ by subtracting the dynamical
phase $\phi^{D}(k,t)$: $\phi^G(k,t)=\phi(k,t)-\phi^{D}(k,t)$.
The total phase $\phi(k,t)$ is the phase factor of Loschmidt amplitude in its polar coordinate representation,
i.e., ${\cal L}_{k}(t)=|{\cal L}_{k}(t)|e^{i\phi(k,t)},$ and $\phi^{D}(k,t)=-\int_0^t \langle \psi_{k}^{f}(t')|H(k,t')|\psi_{k}^{f}(t')\rangle dt',$
in which $\phi(k,t)$ and $\phi^{D}(k,t)$, for the two level system can be calculated as follows \cite{Divakaran2016,Sharma2016b}

%
\bea
\no
\phi(k,t)= \tan^{-1} \Big(\frac{-|u_k|^2 \sin(2\epsilon_k^f t)}{|v_k|^2 + |u_k|^2 \cos (2\epsilon_k^f t)} \Big),~~
\phi^{D}(k,t) = -2 |u_k|^2 \epsilon_k^f t,
\eea
%
%
%
so that \cite{Divakaran2016,Sharma2016b}
%
\bea
\label{eq6}
\phi_k^G = \tan^{-1} \Big(\frac{-|u_k|^2 \sin(2\epsilon_k^f t)}
{|v_k|^2 + |u_k|^2 \cos (2\epsilon_k^f t)} \Big)  + 2 |u_k|^2 \epsilon_k^f t.
\eea
%
In the following we will study the scaling and universality of DQPTs and the corresponding topological properties (DTOP) of the
periodically time dependent extended $XY$ model with linear time dependent transverse field.

\section{Model and Exact Solution\label{model}}
%
\begin{figure*}
\begin{minipage}{\linewidth}
\centerline{\includegraphics[width=0.33\linewidth]{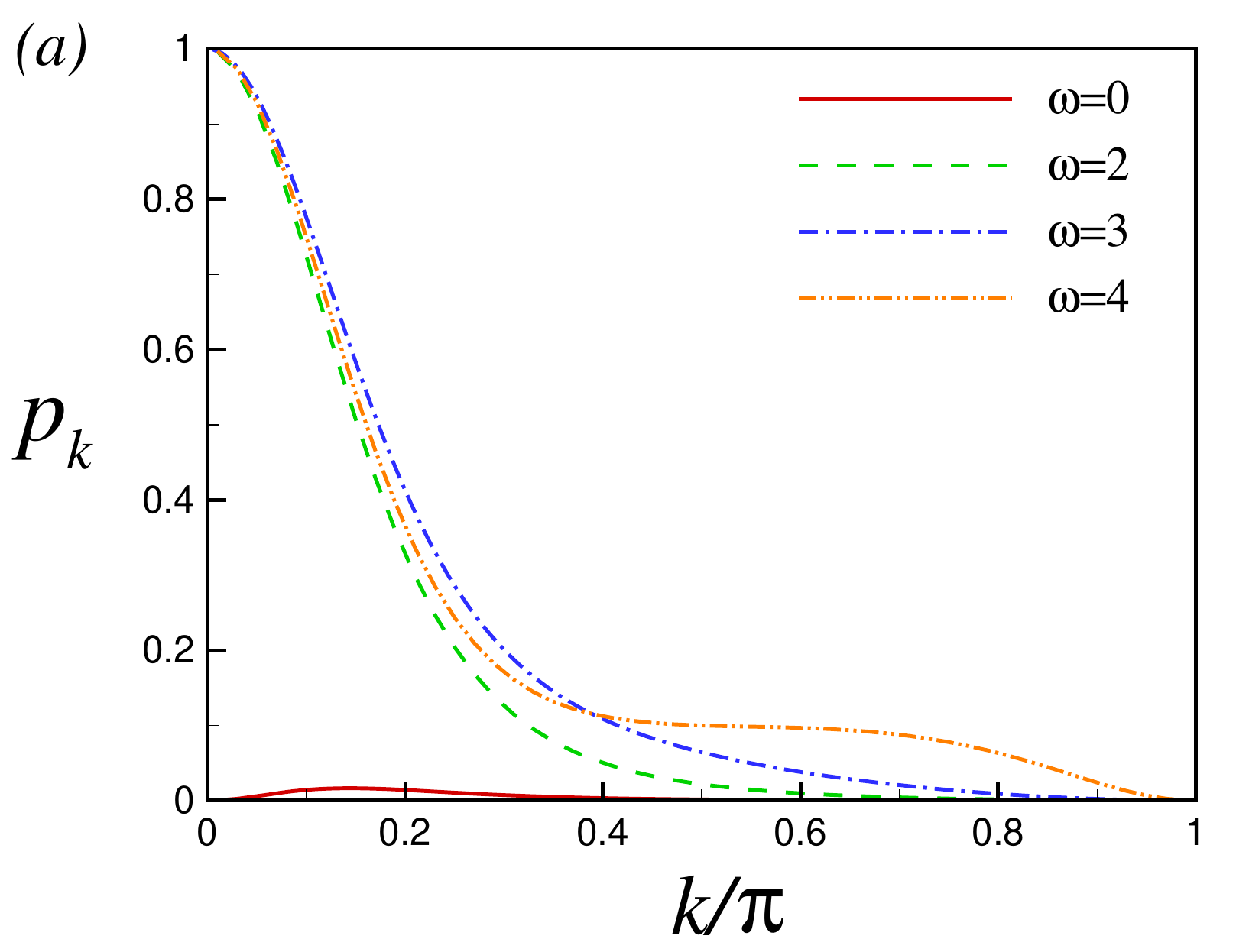}
\includegraphics[width=0.33\linewidth]{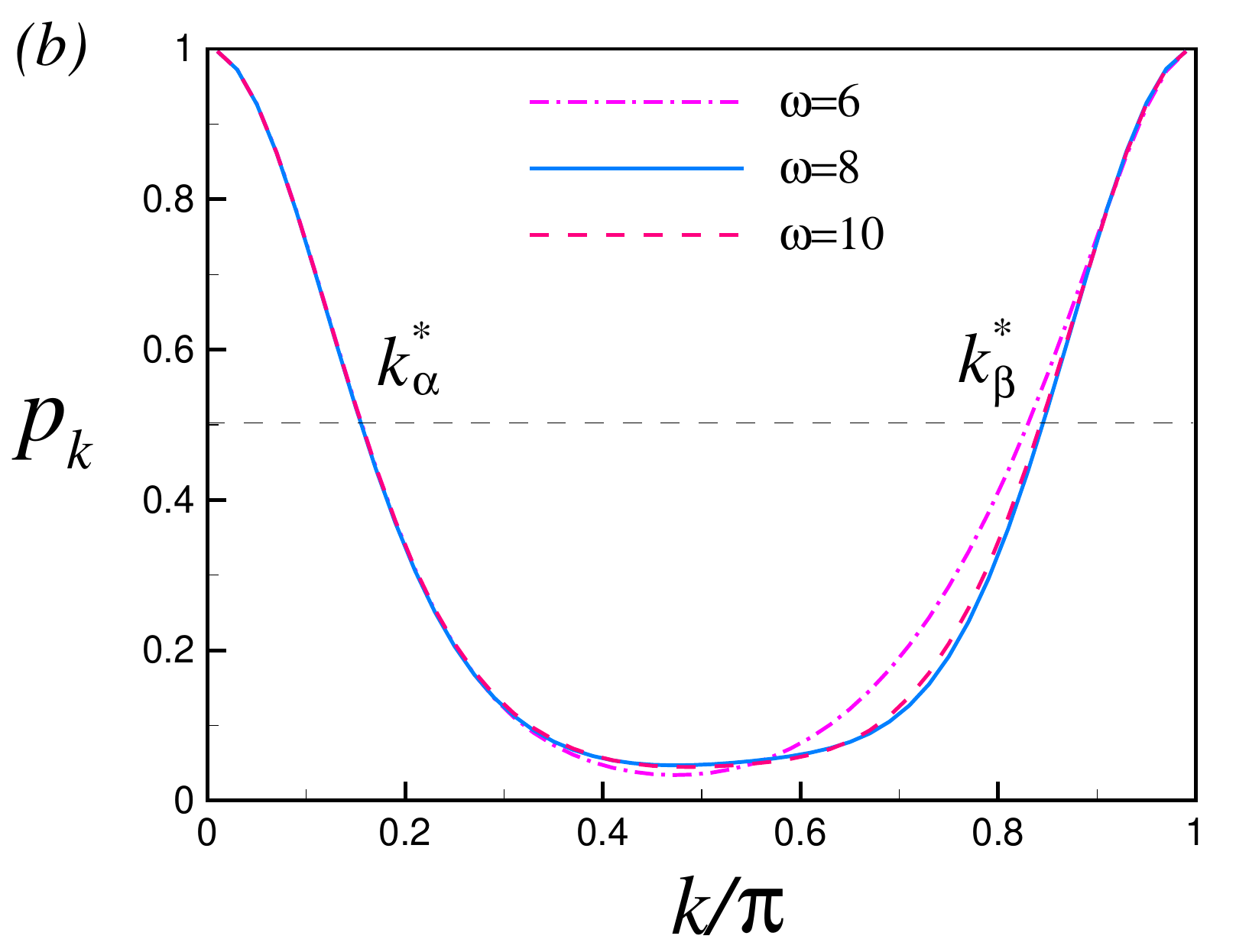}
\includegraphics[width=0.33\linewidth]{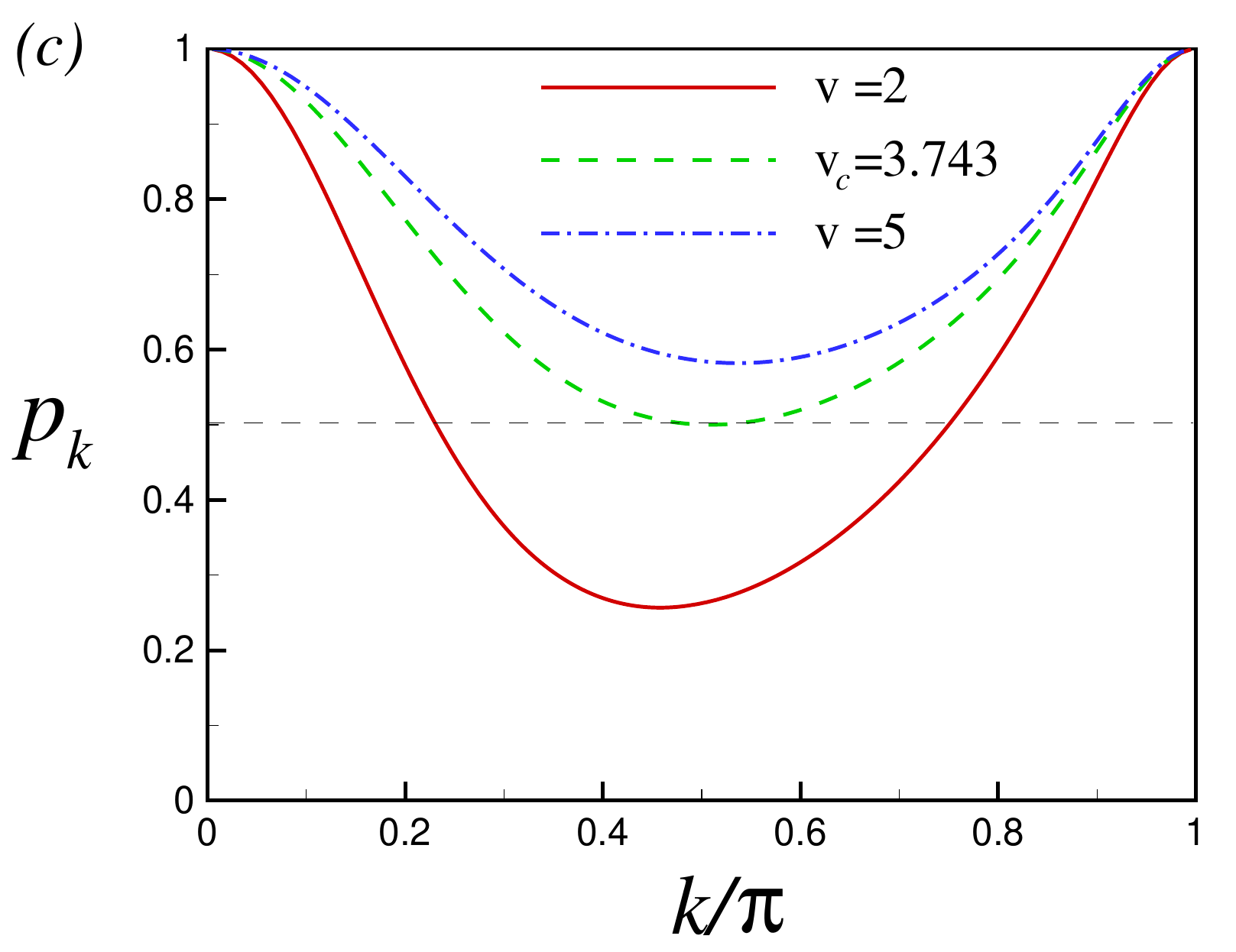}}
\centering
\end{minipage}
\begin{minipage}{\linewidth}
\centerline{\includegraphics[width=0.33\linewidth]{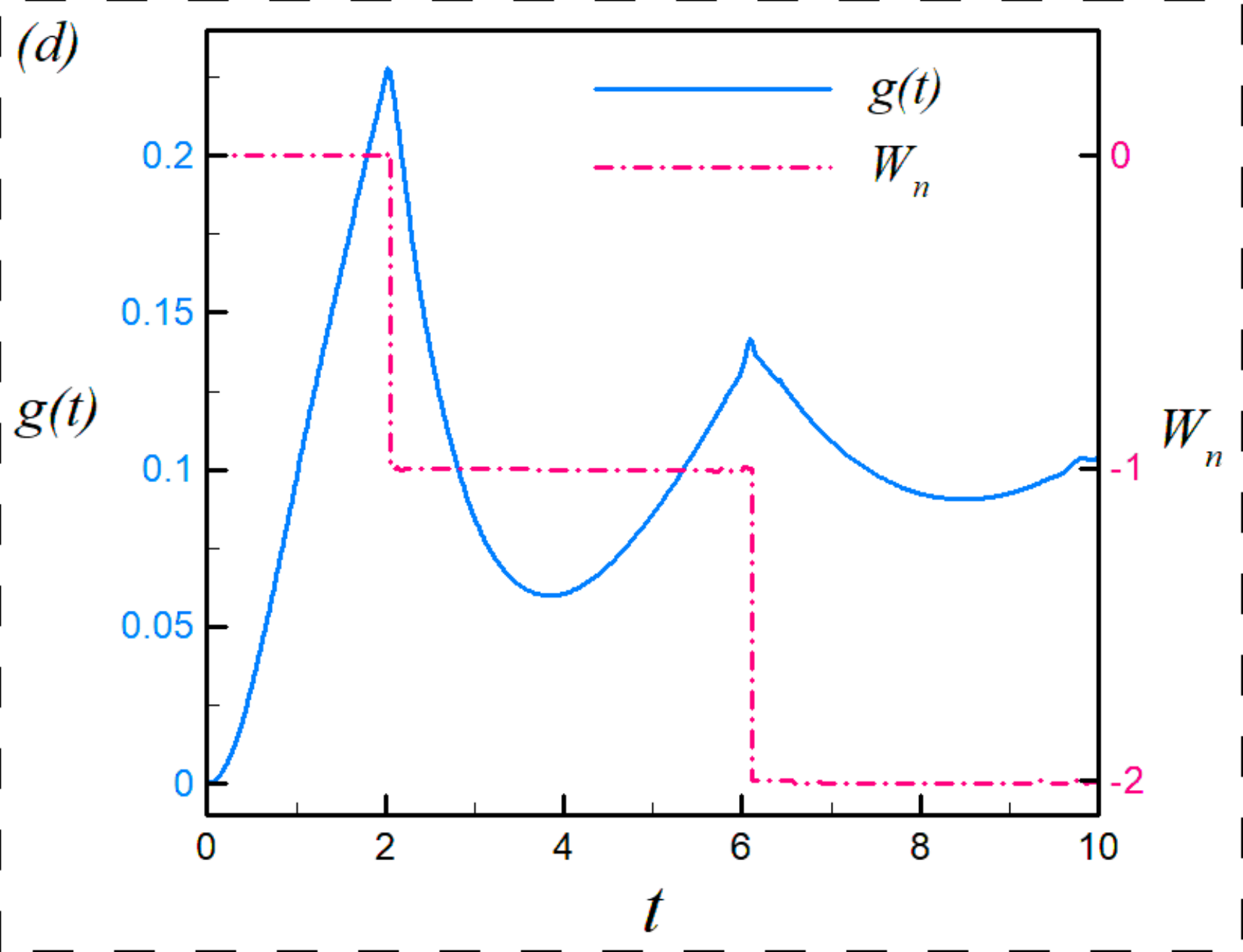}
\includegraphics[width=0.33\linewidth]{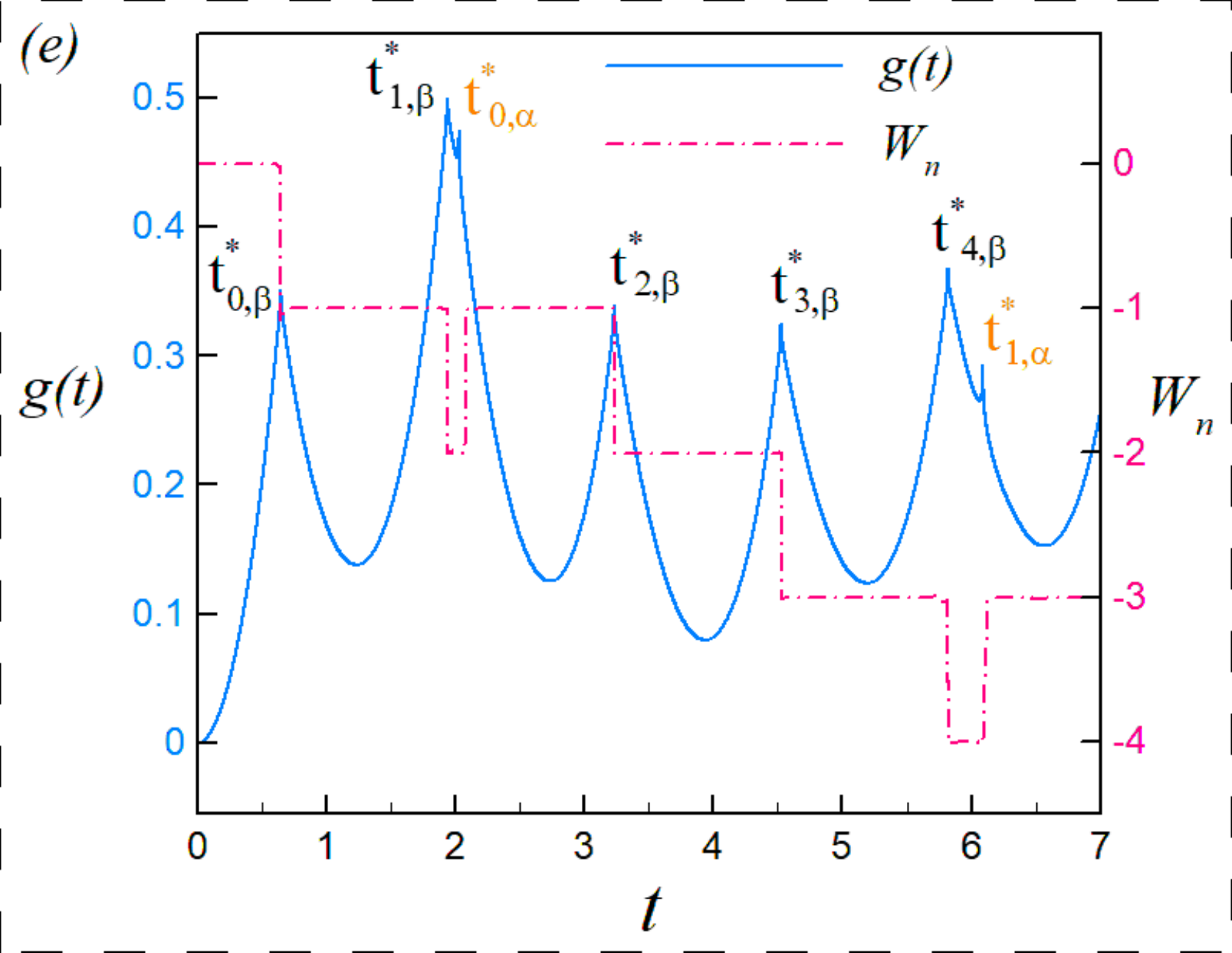}
\includegraphics[width=0.33\linewidth]{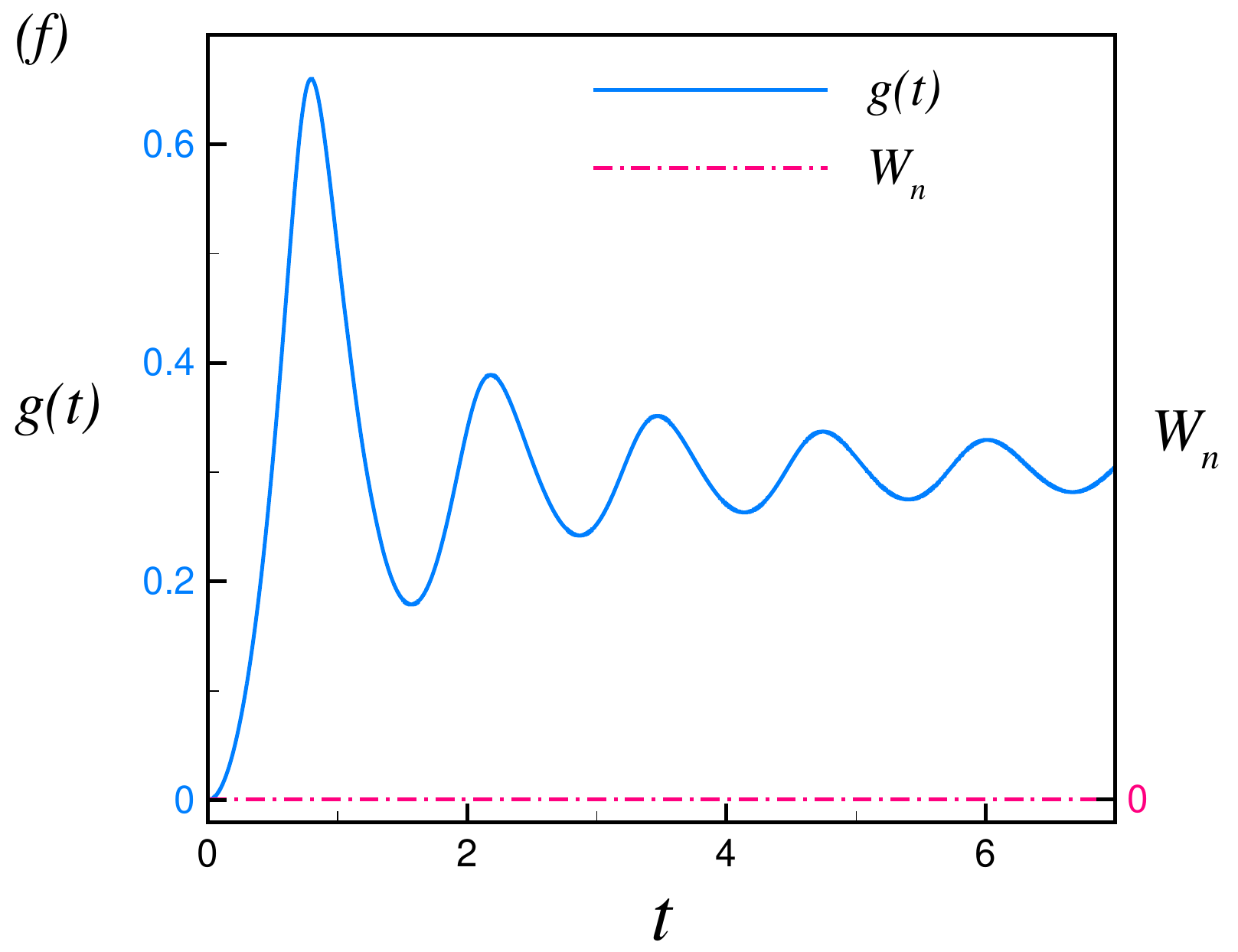}}
\centering
\end{minipage}
\caption{(Color online) Probabilities $p_k$ for finding the system with
momentum $k$ in the upper level for sweep velocity $v=1$, (a) for driven frequency $\omega=0$
at which the ramp quench does not cross the critical point and also for the ramp across the
single critical point $h_c=-1-\omega/2$ for $\omega=2,3,4$, (b)  for the ramp quench that crosses the
two critical points $h_c=-1-\omega/2$ and $h_c=1-\omega/2$ for $\omega=6,8,10$.
(c) The excitation probability versus the momentum for a quench across two critical points $h_c=-4$ and $h_c=-2$ ($\omega=6$)
for different values of the sweep velocity.
The dynamical free energy $g(t)$ and associated dynamical topological order parameter $W_n(t)$  (d) for a quench across the single
critical point $h_c=-2$ ($\omega=2$) for $v=1$, (e) for a quench across two single critical points $h_c=-5$ and $h_c=-3$ ($\omega=8$) for $v=1$,
and (f) for a quench across two critical points $h_c=-4$ and $h_c=-2$ ($\omega=6$) for $v=5>v_c=3.743$.}
\label{fig2}
\end{figure*}
%
The Hamiltonian of Floquet spin system in the presence of linear time dependent magnetic field  is given as
%
\bea
\label{eq7}
{\cal H}(t)=\sum_{n}\Big[(J+\gamma\cos\omega t)s_{n}^{x}s_{n+1}^{x}+(J-\gamma\cos\omega t)s_{n}^{y}s_{n+1}^{y}
-\gamma\sin\omega t(s_{n}^{x}s_{n+1}^{y}+s_{n}^{y}s_{n+1}^{x})+h(t) s_{n}^{z}\Big],
\eea
%
where $s^{\alpha}_{n}=\sigma_{n}^{\alpha}/2$ ($\alpha=x,y,z$) and $\sigma_{n}^{\alpha}$ are the Pauli matrices,
and the transverse magnetic field $h(t)=h_f+vt$ changes from the initial value $h_i$ at time $t=t_i<0$ to the final values $h_f$ at $t=t_f=0$
with sweep velocity $v$.

Performing the Jordan-Wigner fermionization and thanks to the Fourier transformation \cite{Jafari2012}
the Hamiltonian of Eq. (\ref{eq1}) can be written as the sum of $N/2$ non-interacting terms
%
\bea
\label{eq8}
H(t) = \sum_{k} {\cal H}_{k}(t).
\eea
%
with
%
\bea
\label{eq9}
{\cal H}_{k}(t)=[J\cos(k)+h(t)](c_{k}^{\dagger}c_{k}+c_{-k}^{\dagger}c_{-k})
+{\it i}\gamma\sin(k)(e^{{\it i}\omega t}c_{k}^{\dagger}c_{-k}^{\dagger}+e^{-{\it i}\omega t}c_{k}c_{-k}),
\eea
%
where the wave number $k$ is equal to $k=(2m-1)\pi/N$ and $m$ runs from $1$ to $N/2$, being $N$ the total number of spins (sites) in the chain.
Eq. (\ref{eq9}) implies that the Hamiltonian of $N$ interacting spins (Eq. (\ref{eq7})) can be mapped to the
sum of $N/2$ noninteracting quasi-spins.\\

The Bloch single particle Hamiltonian ${\cal H}_{k}(t)$ can be written as
%
\bea
\label{eq10}
{\cal H}_{k}(t)=
\left(
\begin{array}{cc}
h_{z}(k,t) & {\it i}h_{xy}(k)e^{{\it i} \omega t} \\
-{\it i}h_{xy}(k)e^{-{\it i} \omega t}  & -h_{z}(k,t) \\
\end{array}
\right),
\eea
%
with $h_{xy}(k) =\gamma\sin(k)$, and $h_{z}(k,t)=J\cos(k)+h(t)$, which is exactly the same as twisted Landau-Zener Hamiltonian \cite{Takahashi2022,Berry1990,Bouwmeester1996,Bouwmeestert1996b}.
Due to the explicit time dependence in Hamiltonian Eq.(\ref{eq7}), the instantaneous eigenvalues and eigenvectors are given by
%
\bea
\label{eq11}
\varepsilon^{\pm}_{k}=\pm\varepsilon_{k}=\pm\sqrt{h^{2}_{z}(k,t)+h^{2}_{xy}(k)},~~~~~~~~~~~~~~~~~~~~~~~~~~~~~~~~~~~~~~~~~~\\
\no
|\chi^{-}_{k}(t)\rangle=e^{{\it i}\omega t}\cos(\frac{\theta_k}{2}(t))|\uparrow\rangle-{\it i}\sin(\frac{\theta_k}{2}(t))|\downarrow\rangle,~~
|\chi^{+}_{k}(t)\rangle=-{\it i}\sin(\frac{\theta_k}{2}(t))|\uparrow\rangle+e^{-{\it i}\omega t}\cos(\frac{\theta_k}{2}(t))|\downarrow\rangle,
\eea
%
where, $$\cos(\frac{\theta_k(t)}{2})=\frac{(\varepsilon_{k}-h_z(k,t))}{\sqrt{2\varepsilon_{k}(\varepsilon_{k}-h_z(k,t))}},~~ \sin(\frac{\theta_k(t)}{2})=\frac{h_{xy}}{\sqrt{2\varepsilon_{k}(\varepsilon_{k}-h_z(k,t))}},$$ and $|\chi^{\pm}_{k,t}\rangle$ 
are the adiabatic basis of the system.

If the system is prepared in its ground state at $t_i\rightarrow-\infty$ ($h_i\ll h_c=-1$), 
($a_1 (t_i)=1$, $a_2 (t_i)=0$ as explained in Appendix {\ref{APB}}) 
the probability that the $k$:th mode is found in the upper level at $t$ is given as 

%
\bea
\label{eq12}
p_k=|U_{22}\cos(\upvartheta_k(t)/2)+U_{12}\sin(\upvartheta_k(t)/2)|^2,
\eea
%
with,
%
\bea
\no
U_{12}(z)=\frac{\Gamma(1-\nu)}{\alpha\sqrt{\pi}}e^{{\it i}\pi/4}\Big[D_{\nu}(z_i)D_{\nu}(-z)-D_{\nu}(-z_i)D_{\nu}(z)\Big],~~
\no
U_{22}(z)=\frac{\Gamma(1-\nu)}{\sqrt{2\pi}}\Big[D_{\nu}(-z_i)D_{\nu-1}(z)+D_{\nu}(z_i)D_{\nu-1}(-z)\Big],
\eea
%
where, $D_{\nu}(z)$ is the parabolic cylinder function  \cite{szego1954,abramowitz1988}, $\Gamma(x)$ is the Euler gamma function \cite{szego1954,abramowitz1988}
$\alpha=\gamma\sin(k)/\sqrt{v}$, $\nu={\it i}\alpha^{2}/2$, $z=e^{-{\it i}\pi/4}\sqrt{2v}\tau_{k}$, $z_i=e^{-{\it i}\pi/4}\sqrt{2v}\tau_{k,i}$,
$\tau_{k,i}=(h_f+vt_i+J\cos(k)+\omega/2)/v$,
and
%
\bea
\label{eq13}
\cos(\frac{\upvartheta_k(t)}{2})=\frac{\tilde{\varepsilon}_{k}-(h_z(k,t)+\omega/2)}{\sqrt{2\tilde{\varepsilon}_{k}
[\tilde{\varepsilon}_{k}-(h_z(k,t)+\omega/2)]}},~~
\sin(\frac{\upvartheta_k(t)}{2})=\frac{h_{xy}}{\sqrt{2\tilde{\varepsilon}_{k}[\tilde{\varepsilon}_{k}-(h_z(k,t)+\omega/2)]}}.
\eea
%
with $\tilde{\varepsilon}_{k}=\sqrt{(h_{z}(k,t)+\omega/2)^{2}+h^{2}_{xy}(k)}$.
It should be mentioned that, when the magnetic field is time-independent $h(t)=h$, the Floquet Hamiltonian
in the rotating frame given by the unitary transformation $U_{R}(t)=\exp[{\it i}\omega(\sigma^{z})t/2]$  \cite{yang2019floquet,Zamani2020,Jafari2021,Naji2022,Jafari2022,Naji2022b},
is transformed to the time-independent Hamiltonian
%
\begin{equation}
\mathbb{H}_{k}=[h_{xy}(k)\sigma^{x}+(h_{z}(k)+\omega/2)\sigma^{z}].
\label{eq14}
\end{equation}
%
It can be verified that the effective time-independent Hamiltonian undergoes quantum phase transitions
at $h_c=\pm J-\omega/2$, where the energy gap closes at $k=0,\pi$ \cite{yang2019floquet,Naji2022b}.
Therefore the critical points of the effective Hamiltonian can be displaced by tuning the driven frequency.

\section{numerical results\label{results}}
In this section, we report the results of our numerical simulations, based on an analytical approach, to
investigate the dynamics of model using the notion of the DQPTs.
To this end, we consider the linear quenching of the transverse field $h(t) = h_f+vt$, changes from initial value $h_i=-20$ at $t_i=(h_i-h_f)/v$,
where the system is prepared in its ground state, to the final values $h_f=-1.5$ at $t_f=0$.
If we set $J$ to unity, the critical points of the model are given as $h_c=-1-\omega/2$ ($k=0$) and  $h_c=1-\omega/2$ ($k=\pi$).

\subsection{Dynamical free energy and DTOP}

\subsubsection{Quench across a single critical point}

For the ramp quench crosses the single critical point $h_c=-1-\omega/2$ at $k=0$, the excitation probability after the ramp quench depends on the value of $k$.
It is obvious that the modes close to gap closing mode $k=0$  are frozen in the process of quenching as the off-diagonal terms in Hamiltonian Eq. (\ref{eq14}) vanish
and hence $p_{k=0}=1$. While away from the gap closing mode the system evolves adiabatically and can be shown $p_{k\rightarrow\pi}\rightarrow 0$.
Given these two limiting cases, continuity of the transition probability as a function of $k$ in the thermodynamic limit, implies that there exists a critical mode
$k^{\ast}$ at which $p_{k^{\ast}}=1/2$ and consequently DQPTs appear.
The transition probability has been plotted versus $k$ in Fig. \ref{fig2}(a) for sweep velocity $v=1$, and for driven frequency $\omega=0$ whith $h_c=-1$ and also for the ramp across the single critical point $h_c=-1-\omega/2$ for $\omega=2,3,4$.
Since the ramp quench does not cross the critical point for $\omega=0$, the excitation probability is very small. However, $p_{k=0}=1$ for the driven frequencies
$\omega=2,3,4$ at which the quench crosses the single critical point and the transition probability is negligible away from the gap closing mode
($k\rightarrow\pi$).
From these observations, it is straightforward to conclude that there is always a critical momentum $k^{\ast}$ and hence those of $t_n^*$, related through Eq.~(\ref{eq4}), which get modified when $v$ is changed. In other words, different values of sweep velocity simply results in a different sequence of DQPTs time. The existence of DQPTs for $v\to\infty$ is expected, since $k^{\ast}$ is always  present for a sudden quench across a single QCP, as reported in Ref. \cite{Heyl2013}.
%
\begin{figure*}
\begin{minipage}{\linewidth}
\centerline{\includegraphics[width=0.33\linewidth]{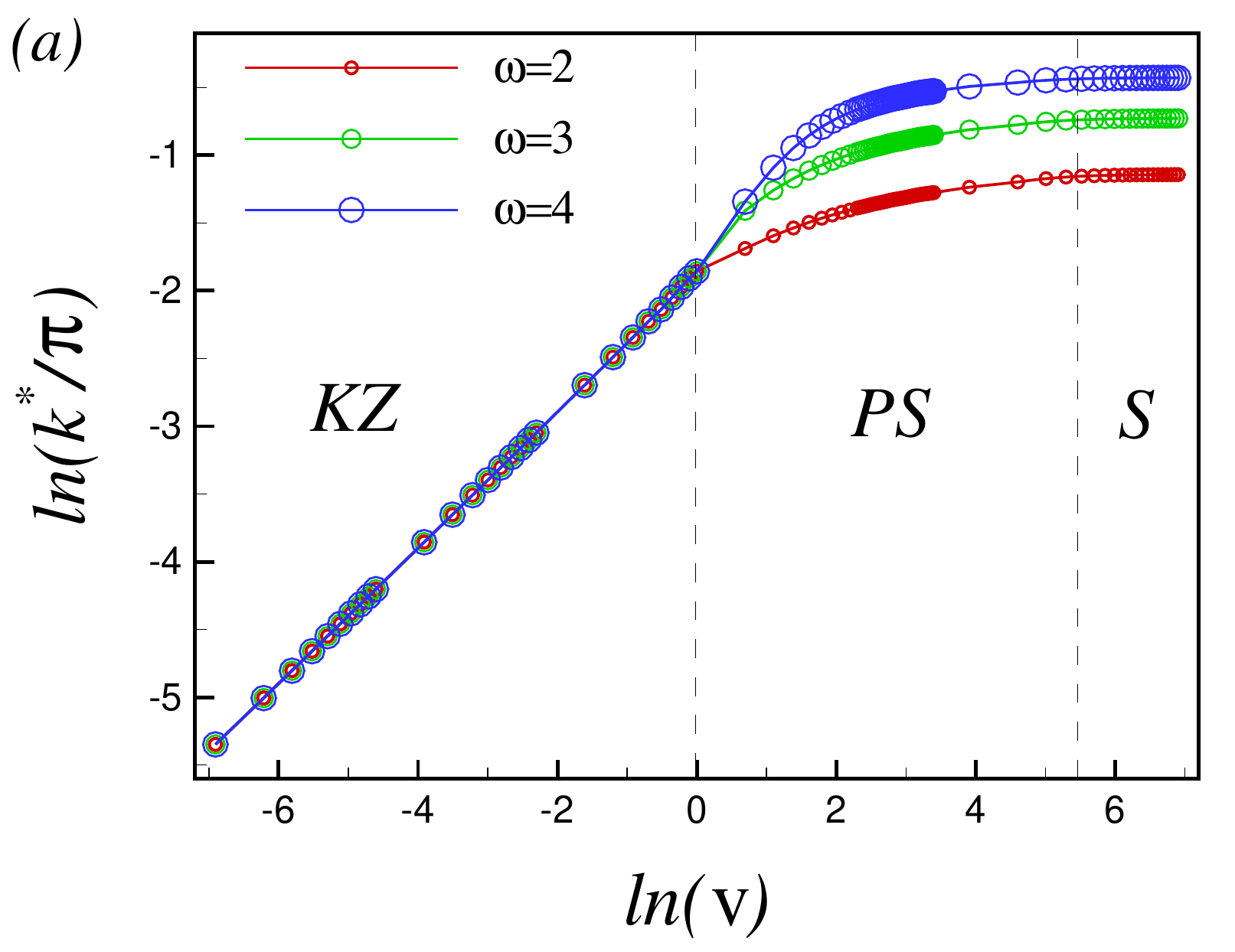}
\includegraphics[width=0.33\linewidth]{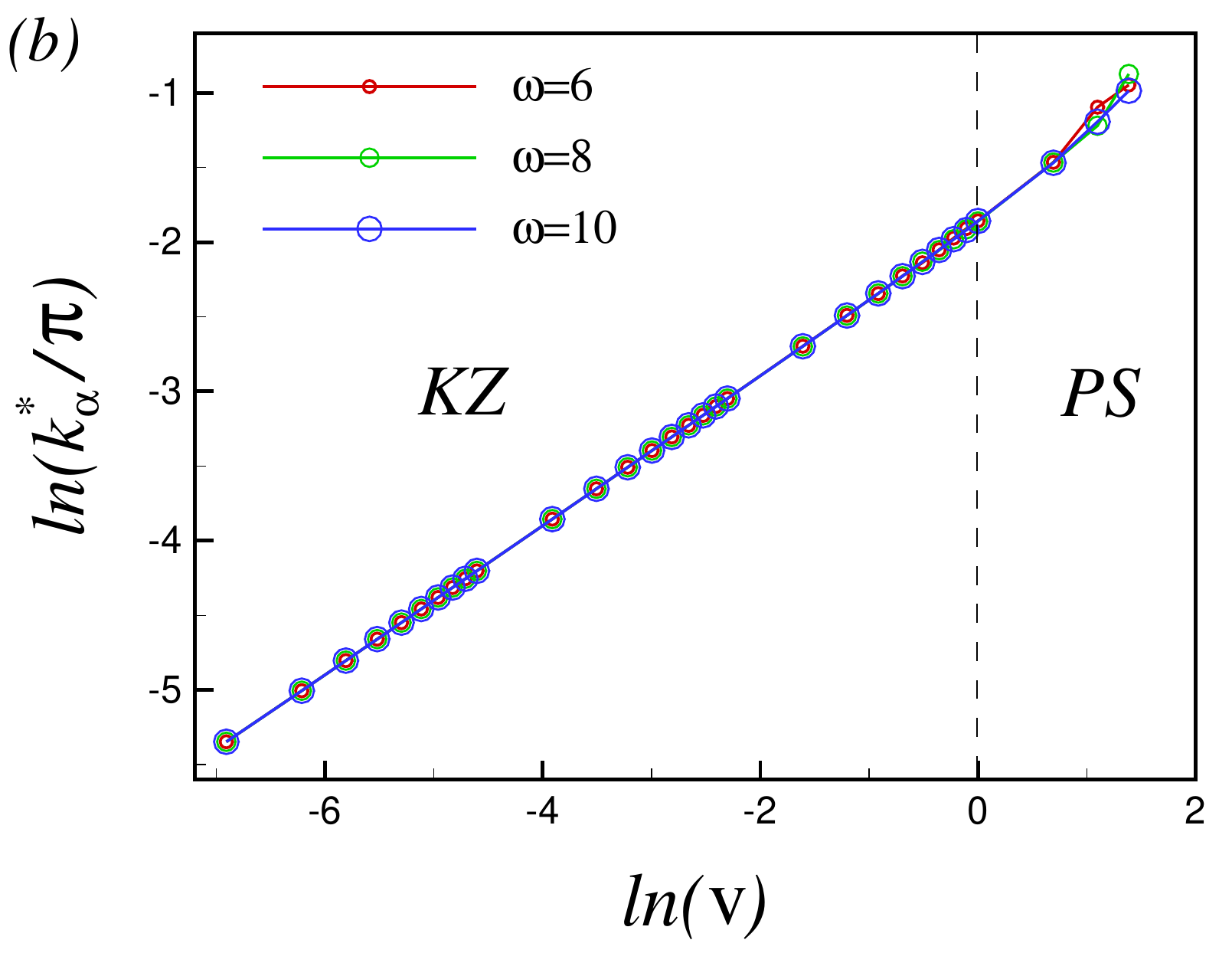}
\includegraphics[width=0.33\linewidth]{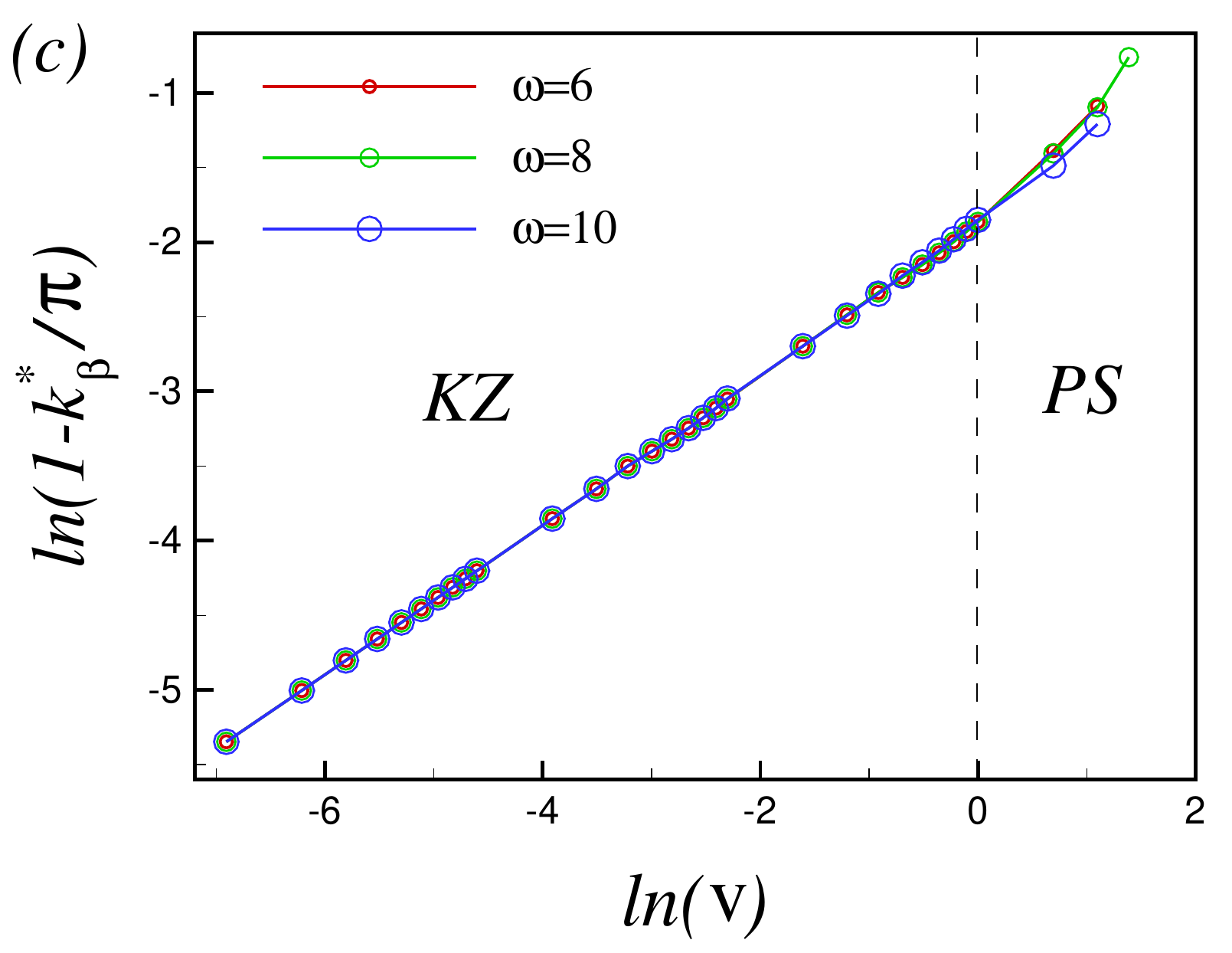}}
\centering
\end{minipage}
\caption{(Color online) Scaling of the critical mode versus sweep velocity for
(a) quench crossing a single critical mode $h_c=-1-\omega/2$ ($\omega=2,3,4$), (b) and (c) quench
across two critical points $h_c=-1-\omega/2$ and $h_c=1-\omega/2$ ($\omega=6,8,10$). 
For a quench that crosses two critical points, there exist
a critical velocity above which DQPT is wiped out. Therefore, 
the scaling behavior of the critical modes include just Kibble-Zurek and pre-saturated regions.}
\label{fig3}
\end{figure*}
%
\subsubsection{Quench across two critical points}
Performing a quench across both equilibrium critical points $h_c=\pm1-\omega/2$ shows new features. Here the field is swept from one equilibrium paramagnetic ground state to
another one, and if the quench is sudden it is not expected to result in DQPTs \cite{vajna2014disentangling,Sharma2016b,Pollmann2013}.  For a quench crossing both critical points, the off-diagonal terms in the Hamiltonian Eq. (\ref{eq14}) are temporally frozen and consequently leading to $p_{k=0}=p_{k=\pi}=1$. As discussed, the maximum value of transition probability $p_{k=0,\pi}=1$  is greater than $1/2$, the occurrence of DQPTs requires the condition that the minimum value of the non-adiabatic transition probability must be less than $1/2$. Thus, making the quench sufficiently slow ($v < v_c$) ensures that the excitation probability for modes away from the gap closing ones is smaller than $1/2$ and sets of a succession of DQPTs.
In Fig. \ref{fig2}(b) the transition probability has been shown versus $k$ for driven frequencies $\omega=6,8,10$ for $v=1$. As predicted, $p_{k=0,\pi}=1$ and the minimum of
$p_k$ away from the critical modes is less than $1/2$ for the small sweep velocity. In such a case, there is two critical modes $k^{\ast}_{\alpha}$ and $k^{\ast}_{\beta}$ at which $p_{k^{\ast}_{\alpha,\beta}}=1/2$ yields a sequence of DQPTs at the corresponding critical times $t^{\ast}_n=t^{\ast}_{n,\alpha}, t^{\ast}_{n,\beta}, n=0,1,\ldots$.

The probability of excitation has been plotted versus $k$ for $\omega=6$ for different values of sweep velocity in Fig. \ref{fig2}(c). As seen there is a threshold sweep velocity $v_c=3.743$ above which the minimum of $p_k$ is greater than $1/2$ and DQPTs get completely wiped out.

The dynamical free energy and DTOP of the model have been depicted in Figs. \ref{fig2}(d)-(f) for two cases of the ramped quench. Fig. \ref{fig2}(d) represents the dynamical free energy
and DTOP for a quench crossing a single critical point $h_c=-2$  ($\omega=2$). Cusps in $g(t)$ and quantizations in the associated DTOP ($W_n(t)$) are clearly visible as an indicator of DQPTs.

The dynamical free energy and DTOP for a quench across two critical points $h_c=-5$ and $h_c=-3$, ($\omega=8$) have been depicted in Fig. \ref{fig2}(e).  The DQPTs show up as cusps in the dynamical free energy $g(t)$ and, more visibly, as steps in the associated DTOP ($W_n(t)$).

Fig. \ref{fig2}(f) displays dynamical free energy and DTOP for a quench that ramp crossing two critical points for $\omega=6$ for sweep velocity $v=5$. As seen, above the critical seep velocity $v=5>v_c=3.743$, a maximally mixed state $p_k=1/2$ does not appear at the end of the quench, thus blocking the appearance of DQPTs.
%
\begin{figure*}
\begin{minipage}{\linewidth}
\centerline{\includegraphics[width=0.33\linewidth]{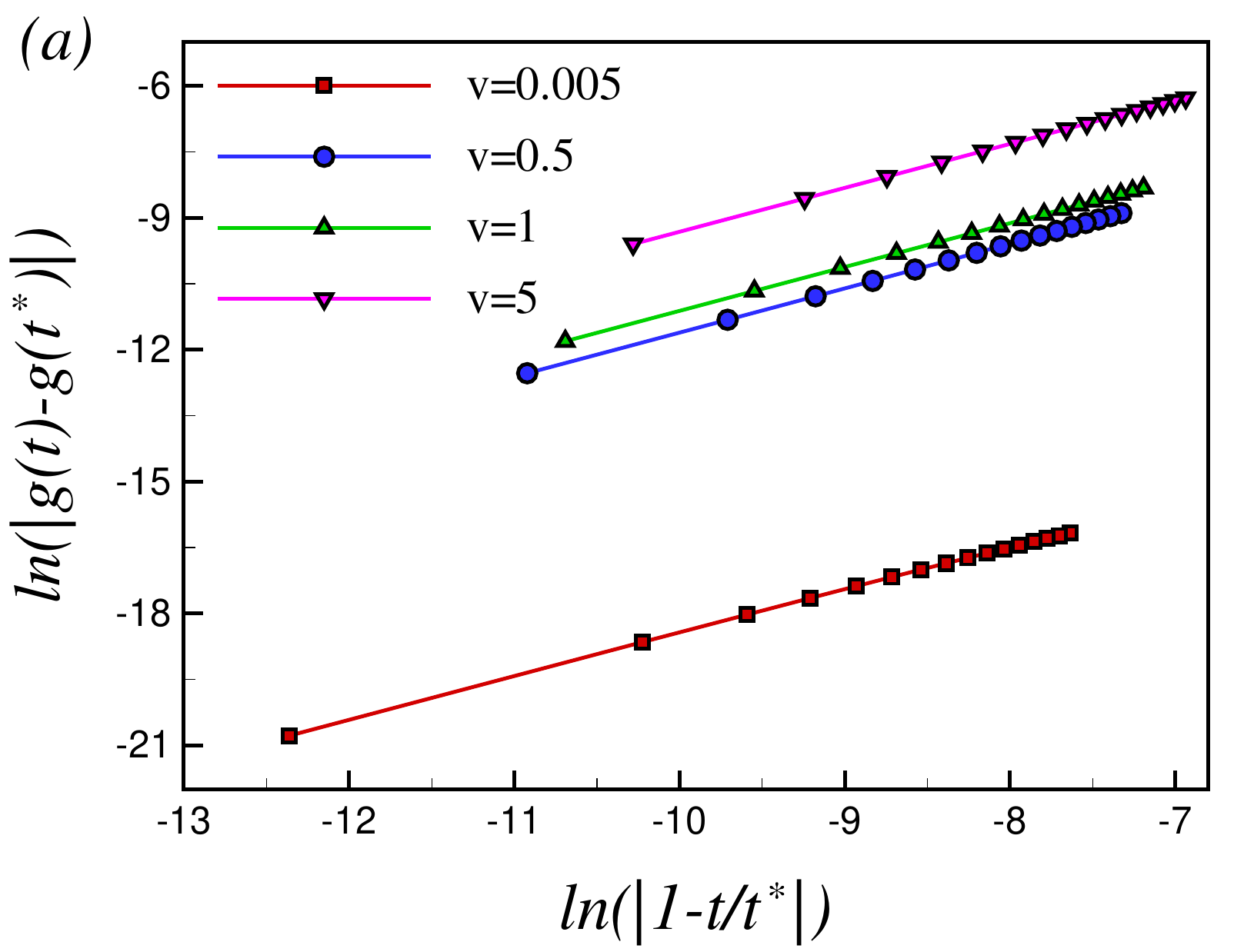}
\includegraphics[width=0.33\linewidth]{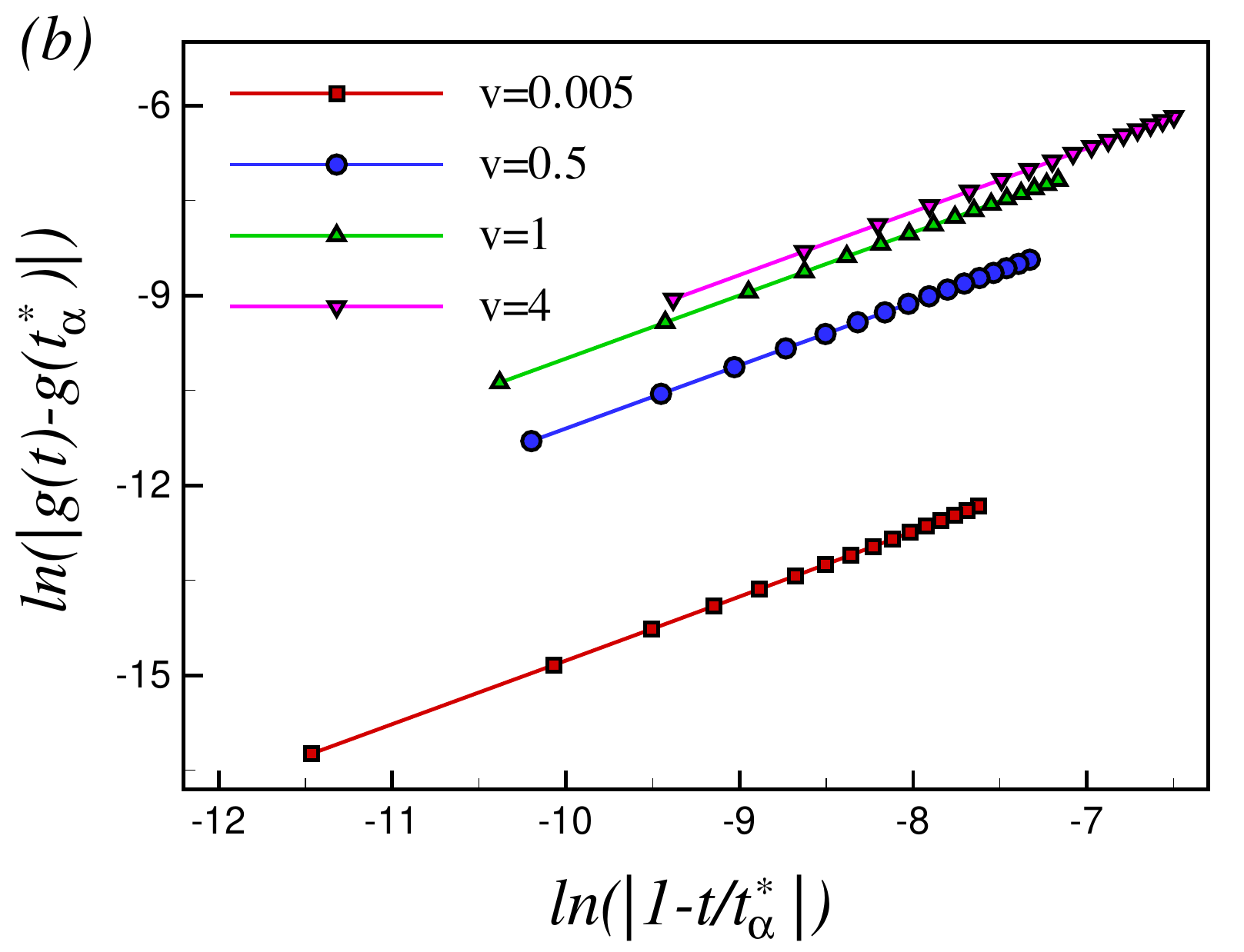}
\includegraphics[width=0.33\linewidth]{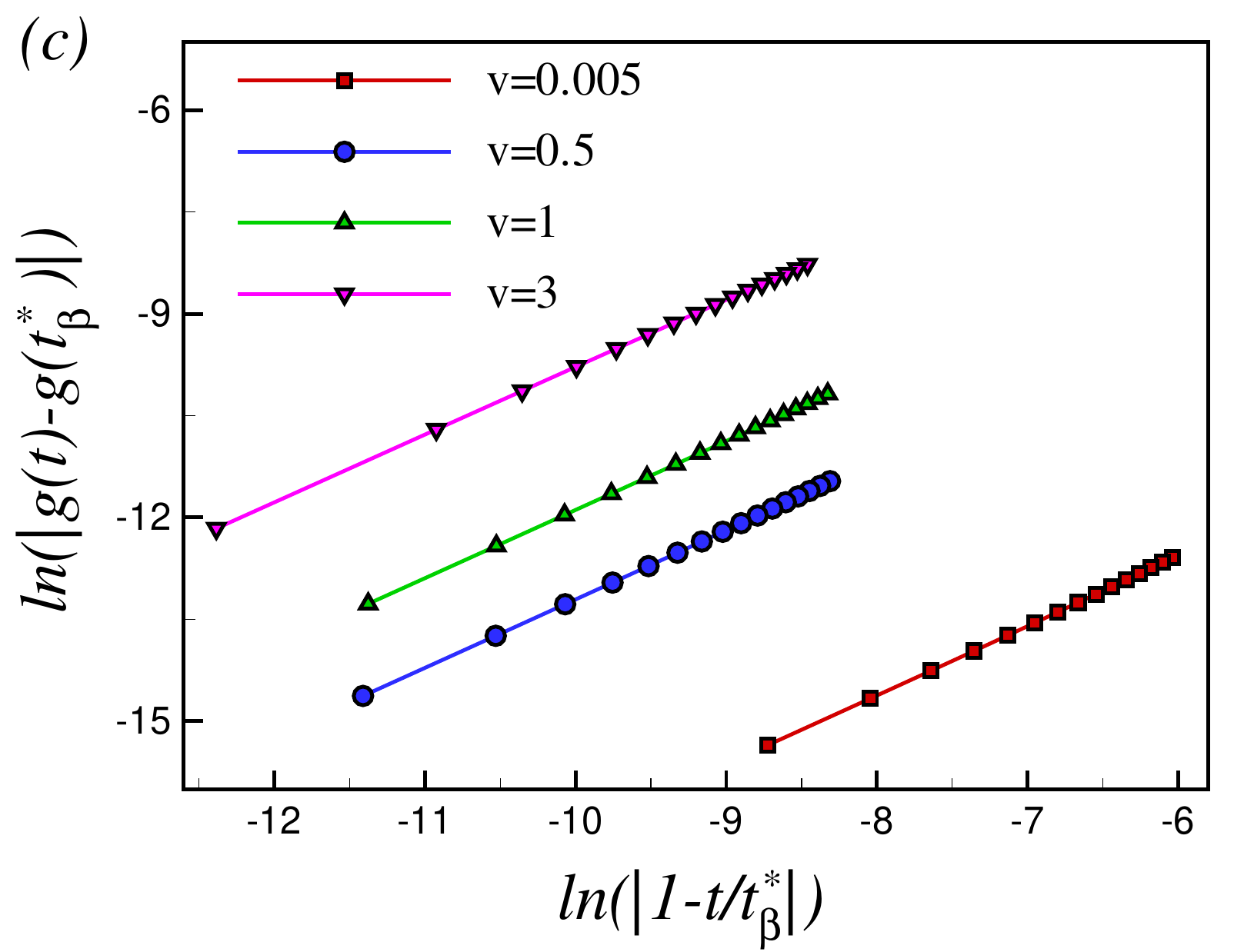}}
\centering
\end{minipage}
\caption{(Color online) . Scaling analysis of the dynamical free energy close to the critical time
for both cases of the ramp quench for different values of the sweep velocity (a) $\omega=2$, 
(b) $\omega=6$ and close to $t^{\ast}_{\alpha}$ and (c) $\omega=6$ and close to $t^{\ast}_{\beta}$.}
\label{fig4}
\end{figure*}
%
 \subsection{Scaling of the critical modes}
Here, we give a comprehensive characterization of the dynamical properties of the ramped quench DQPTs.
To this end, we make the distinction between critical modes of very slow, fast, and very fast ramped quench DQPTs by determining the crossover
points between the three different regions. As a main result, we obtain the behaviour of critical modes as a function of sweep velocity for each region.

\begin{description}
\item[KZ Region] In KZ region the sweep velocity is very small so that we can consider $\tau_{k,t_i}\rightarrow-\infty$.
On the other hand, $\tau_{k,t_f}=(h_f+J\cos(k)+\omega/2)/v\gg\tau_T=\gamma\sin(k)/v$, for $k\ll\pi/4$, 
where $\tau_T$ is the tunnelting time in the adiabatic limit of Landau-Zener model \cite{Mullen1989}.
Therefore, the final time instance is well outside the transition time interval for long wavelength modes and $\tau_{k,t_f}$ can
be extended to $+\infty$ \cite{Dziarmaga2005}. In other words, only long wavelength modes can get excited in slow quench. In such a case, we can use Landau-Zener equation for excitation probability $p_k=\exp{[-\pi(\gamma\sin(k))^2/v]}$  which reduces to $p_k=\exp{(-\pi\gamma^{2} k^2/v)}$ for long wavelength modes.
The critical mode at which the DQPTs happen is given as $k^{\ast}=\sqrt{v \ln(2)/\pi\gamma^2}$. Thus the critical mode scales linearly with the square root of the sweep velocity $v$ in KZ region. It has also been shown that,  in this region the average density of defects excited scales like a square root of the transition rate \cite{Dziarmaga2005}. The dependence of the critical mode on the sweep velocity has been shown in Fig. \ref{fig2}(a)-(c) for two cases of the ramped quench, crossing the single critical point (Fig. \ref{fig3}(a)) and crossing two critical points (Fig. \ref{fig3}(b)-(c)), for different values of the driven frequency. The numerical results show that the critical mode in KZ region collapses on a single curve for all driven frequencies which means that the scaling behaviour of the critical mode is independent of the distance between the final quench field end and the critical point. The numerical analysis confirms that, critical mode scales linearly with the sweep velocity with the exponent $\nu=1/2\pm0.01$.

\item[PS Region] In this region although the quench is fast $v>1$ but not so fast to have $\tau_{k,t_i}\ll 1$. One can ensure this situation
by preparing the initial system far from the
critical point $|h_i|\gg |h_c|$  and also $|h_i+\omega/2|/\sqrt{v}>1$.  Moreover, $\tau_{k,f}$ falls into the tunneling interval time so that we can consider $\tau_{k,t_f}\rightarrow 0$.
As seen in Fig. \ref{fig3}(a)-(c), the critical mode curves belong to the different driven frequency start to get split as the sweep velocity gets faster than one ($v>1$), for both cases of the ramped quench and enhances non-linearly with the sweep velocity. For a quench across the single critical point (Fig. \ref{fig3}(a)) the critical mode enhances with sweep velocity till $v\approx |h_i+\omega/2|^2$.  While in the case of quench that crosses two critical points the PS region is confined between the sweep velocity $v\approx1$, below which the system is in KZ region, and the critical sweep velocity $v_c$, above which the DQPT is disappeared (Fig. \ref{fig3}(b)-(c)).

\item[S Region] In the saturated region, which appears just in a quench across the single critical point  (Fig. \ref{fig3}(a)) the quench is so fast that the evolution lasts only for a
short period of time. Hence  both initial and final times are  well inside the tunneling time interval and  both $\tau_{k,t_i}$ and $\tau_{k,t_f}$ are small such that we  can consider $\tau_{k,t_i},\tau_{k,t_f}\rightarrow 0$, which requires $|h_{i,f}+\omega/2|/\sqrt{v}\ll 1$. In this region the critical mode makes a plateau versus the sweep velocity. This region shrinks with $|h_i|$ enhancing
until it vanishes in the limit $|h_i+\omega/2|\rightarrow\infty$ so that the PS regime prevails the entire fast quench regime.

\end{description}

 \subsection{Scaling of dynamical free energy}
As mentioned, DQPTs at critical times appear as nonanalyticities during nonequilibrium real-time quantum evolution.
In a spirit similar to the equilibrium phase transitions, a major challenge in this work is to seek the fundamental concepts
such as scaling and universality in the ramped quench DQPTs.
It has been proved that the dynamical free energy shows scaling close to the critical time \cite{Heyl2015} at a sudden quench. 
In a one dimensional system the dynamical free energy shows
power law scaling close to the DQPT time
%
\bea
\no
|g(t)-g(t^{\ast})|\sim \Big|\frac{t-t^{\ast}}{t^{\ast}}\Big|^\nu,
\eea
%
with an exponent $\nu=1$, while the scaling law in two dimensional systems is logarithmic \cite{Heyl2015}.
On this basis, we have numerically probed the scaling of the dynamical free energy for two cases of the ramped quench for different values of
sweep velocity and driven frequency. The scaling behaviour of dynamical free energy close to the critical time $t^{\ast}$ has been plotted in Fig. \ref{fig4}(a)
for a quench across the single critical point $h_c=-1$ ($\omega=2$), and for a quench that crosses two critical points $h_c=-4$ and $h_c=-2$ ($\omega=6$)
in Fig. \ref{fig4}(b)-(c) for different values of sweep velocity. Our numerical simulation reveals that the  dynamical free energy shows power law scaling close to the
critical time with the exponent $\nu=1 \pm 0.01$ which shows the same behavior as given in the sudden quench.

\section{Summary and discussion\label{SD}}
In this paper, we have studied the non-equilibrium dynamics of the periodically driven extended $XY$ model in the presence of the
linear time dependent magnetic field. We show that the critical points of the model can be controlled by the driven frequency.
We take advantage of this property to examine the nonequilibrium dynamics of the model after a ramped quench of the magnetic filed.
For a quench across one of the equilibrium quantum critical points, we find that the critical mode at which DQPTs happens, can be classified into
three regions, Kibble-Zurek, pre-saturated and saturated regions. In the Kibble-Zurek region, where the sweep velocity is very small ($v<1$), the critical mode scales
linearly with the square root of the sweep velocity. In this region all curves of the critical mode for different values of the driven frequency collapse on a single
curve which indicates that the scaling properties is independent of the distance between the equilibrium phase transition point and the final magnetic field.
In the pre-saturated region where sweep velocity redirected between $v>1$ and square of the initial magnetic field ($v<|h_i+\omega/2|^2$), the curves of the critical mode for different values of the
driven frequency are splitted and the critical mode does not show any specific scaling.
In the saturated region $v>|h_i+\omega/2|^2$, although curves of the critical mode for different driven frequency are splitted but all curves make a plateau versus the sweep velocity.
The saturated region shrinks by increasing the initial magnetic field and disappear in the limit $|h_i+\omega/2|\rightarrow\infty$ so that the PS region dominates the entire fast quench region.
For a quench that crosses two critical points, there exist a critical velocity above which DQPT is wiped out. In the latter case, the scaling behavior of the critical modes include just Kibble-Zurek and pre-saturated regions.
In addition, our numerical simulations have shown that the dynamical free energy close to the DQPTs time shows power law scaling with the exponent $\nu=1\pm0.01$
for all sweep velocities and driven frequencies, which behaves the same as the sudden quench scaling features of the one dimensional systems.

\appendix

\section{Jordan-Wigner transformation \label{APA}}
The Jordan-Wigner fermionization is defined by following relations
%
\bea
\no
s^{+}_{n}= s^{x}_{n} + {\it i}s^{y}_{n}=\prod_{m=1}^{n-1}(1-2c_{m}^{\dagger}c_{m})c_{n}^{\dagger},~~
s^{-}_{n}= s^{x}_{n} - {\it i}s^{y}_{n} = \prod_{m=1}^{n-1}c_{n}(1-2c_{m}^{\dagger}c_{m}),~~
s^{z}_{n}&=&s^{+}_{n}s^{-}_{n}-\frac{1}{2}= 2c_{n}^{\dagger}c_{n} -1.
\eea
%
where $c_{n}^{\dagger}, c_{n}$ are the spinless fermion creation and annihilation
operators, respectively, and applying the Fourier transform
%
\bea
\no
c_{n} = \frac{1}{\sqrt{N}} \sum_{k} c_{k} e^{-{\it i}kn},~~ c_{n}^{\dagger} = \frac{1}{\sqrt{N}} \sum_{k} c_{n}^{\dagger} e^{{\it i}kn} \; ,
\eea
%
the Hamiltonian of Eq. (\ref{eq7}) can be written as the sum of $N/2$ non-interacting terms
%
\bea
\label{eqAPA1}
H(t) = \sum_{k} {\cal H}_{k}(t).
\eea
%

\section{Time-dependent Schr\"{o}dinger equation in the diabatic basis \label{APB}}

The time-dependent Schr\"{o}dinger equation of Hamiltonian in Eq. (\ref{eq8}) is given as
%
\bea
\label{eqAPB1}
{\it i}\frac{d}{dt}
\left(
 \begin{array}{c}
 a_1(t) \\
 a_2(t) \\
\end{array}
\right)
={\cal H}_{k}(t)\left(
 \begin{array}{c}
 a_1(t) \\
 a_2(t) \\
\end{array}
\right).
\eea
%
To solve  the time-dependent Schr\"{o}dinger equation, we first use the transformation
%
\begin{eqnarray}
\label{eqAPB2}
\left(
               \begin{array}{c}
                a_1(t) \\
                 a_2(t)\\
               \end{array}
\right)
=S
 \left(
               \begin{array}{c}
                 \tilde{a}_1(t) \\
                 \tilde{a}_2(t)\\
               \end{array}
 \right),
\end{eqnarray}
%
with
%
\begin{eqnarray}
\label{eqAPB3}
 S=
  \left(
           \begin{array}{cc}
             0 & {\it i}e^{{\it i}\omega t/2} e^{-{\it i}\pi/4}\\
             -{\it i}e^{-{\it i}\omega t/2} e^{{\it i}\pi/4} & 0 \\
           \end{array}
 \right),
 \end{eqnarray}
%
which results in
%
{\small
\bea
\no
{\it i}\frac{d}{dt}
\left(
 \begin{array}{c}
 \tilde{a}_2(t) \\
 \tilde{a}_1(t) \\
\end{array}
\right)
=
\left(
\begin{array}{cc}
 -(h_z(k,t)+\frac{\omega}{2}) & \gamma\sin(k)  \\
\gamma\sin(k) &h_z(k,t)+\frac{\omega}{2}\\
\end{array}
\right)
\left(
 \begin{array}{c}
 \tilde{a}_2(t) \\
 \tilde{a}_1(t) \\
\end{array}
\right).\\
\label{eqAPB4}
\eea
}
%
The time-dependent Schr\"{o}dinger equation (\ref{eq10}) is mapped to the time-dependent Schr\"{o}dinger equation
of Landau-Zener problem \cite{Vitanov1996,Vitanov1999} by defining the new time scale $\tau_{k}=(h_f+vt+J\cos(k)+\omega/2)/v$,
%
{\small
\bea
\label{eqAPB5}
{\it i}\frac{d}{d\tau_{k}}
\left(
 \begin{array}{c}
 \tilde{a}_2(\tau_{k}) \\
 \tilde{a}_1(\tau_{k}) \\
\end{array}
\right)
=
\left(
\begin{array}{cc}
 -v\tau_{k} & \gamma\sin(k)  \\
\gamma\sin(k) & v\tau_{k} \\
\end{array}
\right)
\left(
 \begin{array}{c}
 \tilde{a}_2(\tau_{k}) \\
 \tilde{a}_1(\tau_{k}) \\
\end{array}
\right).
\eea
}
%

The Landau-Zener problem is exactly solvable as explained in Refs. \cite{Vitanov1999,Vitanov1996}
%
\bea
\label{eqAPB6}
\left(
 \begin{array}{c}
 \tilde{a}_2(\tau_{k}) \\
 \tilde{a}_1(\tau_{k}) \\
\end{array}
\right)
=
\left(
\begin{array}{cc}
U_{11} & U_{12} \\
U_{21} & U_{22} \\
\end{array}
\right)
\left(
 \begin{array}{c}
 \tilde{a}_2(\tau_{k,i}) \\
 \tilde{a}_1(\tau_{k,i}) \\
\end{array}
\right),
\eea
%
where,
%
\bea
\no
U_{11}(z)&=&\frac{\Gamma(1-\nu)}{\sqrt{2\pi}}\Big[D_{\nu-1}(-z_i)D_{\nu}(z)+D_{\nu-1}(z_i)D_{\nu}(-z)\Big],~~
U_{12}(z)=\frac{\Gamma(1-\nu)}{\alpha\sqrt{\pi}}e^{{\it i}\pi/4}\Big[D_{\nu}(z_i)D_{\nu}(-z)-D_{\nu}(-z_i)D_{\nu}(z)\Big],\\
\no
U_{21}(z)&=&\frac{\alpha\Gamma(1-\nu)}{2\sqrt{\pi}}e^{-{\it i}\pi/4}\Big[D_{\nu-1}(z_i)D_{\nu-1}(-z)-D_{\nu-1}(-z_i)D_{\nu-1}(z)\Big],\\
\label{eqAPB7}
U_{22}(z)&=&\frac{\Gamma(1-\nu)}{\sqrt{2\pi}}\Big[D_{\nu}(-z_i)D_{\nu-1}(z)+D_{\nu}(z_i)D_{\nu-1}(-z)\Big],
\eea
%
where, $D_{\nu}(z)$ is the parabolic cylinder function \cite{szego1954,abramowitz1988},
$\alpha=\gamma\sin(k)/\sqrt{v}$, $\nu={\it i}\alpha^{2}/2$, $z=e^{-{\it i}\pi/4}\sqrt{2v}\tau_{k}$, $z_i=e^{-{\it i}\pi/4}\sqrt{2v}\tau_{k,i}$  and
$\tau_{k,i}=(h_f+vt_i+J\cos(k)+\omega/2)/v$.\\

Using the transformation given in Eq. (\ref{eqAPB2}), we can transform $\tilde{a}_{m}(\tau_{k})$ to $a_{m}(t)$ ($ m=1,2$)
%
\bea
\no
S(t)\left(
 \begin{array}{c}
 \tilde{a}_2(\tau_{k}) \\
\tilde{a}_1(\tau_{k}) \\
\end{array}
\right)
=
S(t)\left(
\begin{array}{cc}
U_{11} & U_{12} \\
U_{21} & U_{22} \\
\end{array}
\right)
S^{\dag}(t_i)S(t_i)
\left(
 \begin{array}{c}
 \tilde{a}_2(\tau_{k,i}) \\
 \tilde{a}_1(\tau_{k,i}) \\
\end{array}
\right),
\eea
%
%
\bea
\no
\left(
 \begin{array}{c}
 a_1(t) \\
a_2(t) \\
\end{array}
\right)
=
S(t)\left(
\begin{array}{cc}
U_{11} & U_{12} \\
U_{21} & U_{22} \\
\end{array}
\right)
S^{\dag}(t_i)
\left(
 \begin{array}{c}
a_1(t_i) \\
a_2(t_i) \\
\end{array}
\right).
\eea
%

\section{Exact solution in the adiabatic basis}
The adiabatic solution can be obtained by transforming Eqs. (\ref{eqAPB1}) into the adiabatic representation by the unitary transformation
%
{\small
\begin{eqnarray}
\label{eqAPC1}
R=
 \left(
           \begin{array}{cc}
             e^{{\it i}\omega t/2}\sin(\upvartheta_k(t)/2) & {\it i} e^{{\it i}\omega t/2}\cos(\upvartheta_k(t)/2)\\
              {\it i} e^{-{\it i}\omega t/2}\cos(\upvartheta_k(t)/2)&  e^{-{\it i}\omega t/2}\sin(\upvartheta_k(t)/2)\\
           \end{array}
 \right)
 \end{eqnarray}
 }
%
where, $$\cos(\frac{\upvartheta_k(t)}{2})=\frac{\tilde{\varepsilon}_{k}-(h_z(k,t)+\omega/2)}{\sqrt{2\tilde{\varepsilon}_{k}
[\tilde{\varepsilon}_{k}-(h_z(k,t)+\omega/2)]}},~~ \sin(\frac{\upvartheta_k(t)}{2})=\frac{h_{xy}}{\sqrt{2\tilde{\varepsilon}_{k}[\tilde{\varepsilon}_{k}-(h_z(k,t)+\omega/2)]}},$$
with $\tilde{\varepsilon}_{k}=\sqrt{(h_{z}(k,t)+\omega/2)^{2}+h^{2}_{xy}(k)}$.
%
\begin{eqnarray}
\label{eqAPC2}
\left(
               \begin{array}{c}
                a^{\mathcal{A}}_1(t) \\
                 a^{\mathcal{A}}_2(t)\\
               \end{array}
\right)=R^{\dagger}
 \left(
               \begin{array}{c}
               a_1 (t)\\
                 a_2(t)\\
               \end{array}
  \right).
 \end{eqnarray}
%
In the limit $\tau_k\gg\gamma\sin(k)$ , the adiabatic eigenstates coincide with the diabatic states.
If the system is prepared in its ground state at $t_i\rightarrow-\infty$ ($h_i\ll h_c=-1$), i,e., $a_1 (t_i)=1$, $a_2 (t_i)=0$
the probability that the $k$:th mode is found in the upper level at $t$ is given as $p_k=|U_{22}\cos(\upvartheta_k(t)/2)+U_{12}\sin(\upvartheta_k(t)/2)|^2$.

\bibliographystyle{iopart-num}
\bibliography{REF}

\end{document}